\documentclass{article}

 \usepackage[final,nonatbib]{neurips_2024}

\usepackage[utf8]{inputenc} % allow utf-8 input
\usepackage[T1]{fontenc}    % use 8-bit T1 fonts
\usepackage{hyperref}       % hyperlinks
\usepackage{url}            % simple URL typesetting
\usepackage{booktabs}       % professional-quality tables
\usepackage{amsfonts}       % blackboard math symbols
\usepackage{nicefrac}       % compact symbols for 1/2, etc.
\usepackage{microtype}      % microtypography
\usepackage{xcolor}         % colors
\usepackage{graphicx} 
\usepackage{amsmath} 
\usepackage{multirow}
\usepackage[normalem]{ulem}

\title{DGFM: Full Body Dance Generation Driven by Music Foundation Models\thanks{\textcolor{red}{This paper was presented at the Audio Imagination Workshop of NeurlPS 2024, but it has not been formally published in any conference proceedings or journals.}}}

\author{
  Xinran Liu$^{1}$, Zhenhua Feng$^{2}$, Diptesh Kanojia$^{1}$, Wenwu Wang$^{1}$ \\
  $^1$University of Surrey, UK, $^2$Jiangnan University, China
}

\begin{document}

\maketitle

\begin{abstract}
  %The use of audio has been a long-standing topic for cross-modal content generation. 
  In music-driven dance motion generation, most existing methods use hand-crafted features and neglect that music foundation models have profoundly impacted cross-modal content generation. To bridge this gap, we propose a diffusion-based method that generates dance movements conditioned on text and music. Our approach extracts music features by combining high-level features obtained by music foundation model with hand-crafted features, thereby enhancing the quality of generated dance sequences. This method effectively leverages the advantages of high-level semantic information and low-level temporal details to improve the model's capability in music feature understanding. To show the merits of the proposed method, we compare it with four music foundation models and two sets of hand-crafted music features. The results demonstrate that our method obtains the most realistic dance sequences and achieves the best match with the input music.
\end{abstract}

\section{Introduction}
Music-driven dance video generation is a challenging task in cross-modal content generation~\cite{li2021ai}. The complexity comes from the need to generate dance movements that follow choreographic rules and are aligned precisely with the music. Specifically, the generated motions must be expressive while corresponding to rhythm, melody, and other musical elements. 
This requires the generative model to have outstanding performance in music feature extraction. 

Early music-driven dance generation approaches 
frame the task as a similarity-based retrieval problem~\cite{aksan2019structured,qiu2019cross,galata2001learning,ofli2011learn2dance} due to limited training data. 
These methods significantly restrict the diversity and creativity of the outputs. 
With the advent of deep learning, recent advancements have treated music-driven dance generation as a generative task~\cite{siyao2022bailando,chen2021choreomaster,li2022danceformer,yang2023longdancediff, huang2022genre}. 
However, most existing models are trained on datasets with limited joint representations (\textit{e.g.}, $24$ joints), neglecting finer hand motion details, which are essential for enhancing the realism and expressiveness of the generated dances. 
Furthermore, the existing approaches rely on hand-crafted musical features such as Mel-Frequency Cepstral Coefficient (MFCC)~\cite{abdul2022mel}, chroma, or one-hot beat features, which cannot fully capture the intricate link between music and dance movements~\cite{qi2023diffdance}. 
In contrast, recent advances in music foundation models have demonstrated the potential of modern machine learning techniques to better understand and process music in more sophisticated ways~\cite{li2023mert}. These models trained on large-scale music datasets with minimal supervision, serve as the foundation for multiple derived models capable of performing a wide range of tasks, including music classification~\cite{huang2024mavil}, music understanding~\cite{cheng2023ssvmr,mckee2023language}, and cross-model generation~\cite{mittal2021symbolic, copet2024simple}.
The potential of music foundation models in music-driven dance generation remains under-explored, making it important to investigate how these models can contribute to the quality and expressiveness of the dance generation task.

In this paper, we propose a diffusion-based method, namely Dance Generation driven by the Music Foundation Models (DGFM), which enhances the quality of the generated dance movements by incorporating the classical hand-crafted audio features with high-level features obtained by the pre-trained large music foundation modelcally, we use Wav2CLIP~\cite{wu2022wav2clip} for high-level music feature extraction. Wav2CLIP is an audio-visual model trained by distilling the knowledge from Contrastive Language-Image Pretraining (CLIP)~\cite{radford2021learning}. Unlike the models that rely solely on text and audio inputs, Wav2CLIP benefits from this additional visual context, enabling more comprehensive learning and significantly improving motion prediction. 
Additionally, we use Short-Time Fourier Transform (STFT) as hand-crafted audio features and employ CLIP to extract features from genre prompts. By building on these components, our method captures a deeper understanding of the relationship between music and movement, resulting in realistic and intricately detailed 3D dance motions.

The contributions of our work can be summarized as follows: (1) We introduce DGFM which integrates both music and text features as inputs. It improves hand-crafted audio features by incorporating Wav2CLIP, significantly enhancing the dance generation quality. (2) To investigate the impact of music understanding on dance motion generation, we compare different music foundation models and hand-crafted music features. The results demonstrate that the combination of Wav2CLIP with STFT features achieves promising results. (3) Extensive experiments performed on the FineDance dataset demonstrate the effectiveness of our approach.\section{Related work}
\label{gen_inst}
\textbf{Music Foundation Model:} Music foundation models are pre-trained on large-scale music datasets, which are designed to gain a deeper understanding of underlying musical structures, genres, and instruments~\cite{ma2024foundation}.
The existing music foundation models can be divided into two categories.
In the first category, the models are pre-trained with a single modality. This category includes Wav2Vec 2.0~\cite{baevski2020wav2vec}, which is a self-supervised model that learns audio representations from raw audio through contrastive learning. Additionally, Li \textit{et al.}~\cite{li2023mert} proposed MERT, a model for music understanding that takes advantage of Residual VQ-VAE (RVQ-VAE) and teacher models to extract musical features, facilitating pre-training based on mask language modeling (MLM). Jukebox~\cite{dhariwal2020jukebox} compresses raw audio into discrete codes using a multi-scale VQ-VAE and models them with an autoregressive Transformer, applied in~\cite{tseng2023edge} to enhance previous hand-crafted audio feature extraction strategies for dance generation task.
In he second category, the models are pre-trained by multi-modal data, such as CLAP~\cite{wu2023large} which leverages the latent space derived from both audio and text to develop continuous audio representations.
Building on this, AudioLDM~\cite{liu2023audioldm} utilizes CLAP to train Latent Diffusion Models (LDMs) with audio embeddings, while text embeddings are used as conditions during sampling. 
Wu \textit{et al.}~\cite{wu2022wav2clip} proposed Wav2CLIP based on CLIP~\cite{radford2021learning}, which projects audio into a shared embedding space along with images and text to pretrain the audio encoder.
However, the application of these models in music-driven dance generation remains under-explored. 
It is important to investigate how music foundation models can contribute to enhancing dance generation tasks.

\textbf{Music Driven Dance Generation:}
There has been extensive research exploring music-conditioned dance generation. 
Early studies~\cite{aksan2019structured,qiu2019cross,galata2001learning} formulate this task as a similarity-based retrieval problem due to limited training data, which substantially limits the diversity and creativity of the generated dance motions. 
With the development of deep learning, the mainstream approaches synthesize dance segments from scratch via motion prediction, with methods convolutional neural network (CNN)~\cite{holden2016deep,holden2015learning}, recurrent neural network (RNN)~\cite{butepage2017deep,chiu2019action,du2019bio}, and Transformer~\cite{huang2022genre,li2020learning,li2022danceformer}.
However, these frame-by-frame prediction models frequently face challenges like error accumulation and motion freezing~\cite{zhuang2022music2dance}.
Recent studies have focused on framing the task as a generative pipeline.
Built on VQ-VAE, TM2D~\cite{gong2023tm2d} integrates both music and text instructions to generate dance movements that are consistent with the provided music and contain semantic information.
Bailando~\cite{siyao2022bailando} summarizes meaningful dance units into a quantized codebook and incorporates a reinforcement-learning-based evaluator to ensure alignment between beats and movement during dance generation. 
EDGE~\cite{tseng2023edge} apply a diffusion-based dance generation model, and also introduce a novel evaluation method based on human physical plausibility. 
However, all these models are trained on datasets with 24 body joints and neglect the quality of the hand motions generated. To mitigate this issue, Li \textit{et al.}~\cite{li2023finedance} proposed FineNet and introduced a new dataset with $52$ joints. 
Despite these developments, almost all the existing models rely on hand-crafted musical features, which may lack an understanding of the connection between music and dance movements.

\section{Proposed Approach}
In this section, we introduce our approach for generating dance movements conditioned on both music sequences and text. 
Given a long music piece, we first split it into $N$ segments with length \( k \) and extract the 2D music feature map $M\in \mathbb{R}^{k \times  D}$, where \( D\) represents the dimension of music features. Our objective is to generate \( N \) dance clips of length \( k \) or a long dance movement. 

The existing approaches usually neglect the importance of music feature representation. Therefore, we present to incorporate Wav2CLIP~\cite{wu2022wav2clip} to our music encoder, which is an audio-visual correspondence model that distills from the CLIP framework. It is trained on VGGSound~\cite{chen2020vggsound}, a YouTube audio-visual video dataset containing approximately $200$k audio clips of length $10$ seconds ($16$kHz sampling rate) labeled with $309$ classes. For hand-crafted music features, we utilize the Librosa toolbox~\cite{mcfee2015librosa} to extract STFT features. 
Additionally, for music genre labeling, we apply a prompt learning method~\cite{zhou2022learning} to expand the label into a full sentence. For \textit{e.g.}, given the genre label "Jazz," the sentence generated is  "This is a Jazz type of music." We then use CLIP~\cite{radford2021learning} to extract features from this sentence.

\subsection{Preliminaries of Diffusion Models}
We use a diffusion model for dance generation, which has two stages: the diffusion process and the reverse process. 
In the diffusion process, we follow the approach outlined in Denoising Diffusion Probabilistic Models (DDPM)~\cite{ho2020denoising}. 
It defines a Markov chain that progressively adds Gaussian noise to the ground truth data $\textbf{\textit{m}}_{0}$, while allowing for the sampling of $\textbf{\textit{d}}_{t}$ at any arbitrary timestep $\mathit{t}$:
\begin{equation}  
q\left(\textbf{\textit{d}}_{t} \mid \textbf{\textit{m}}_{0}\right)=\mathcal{N}\left(\textbf{\textit{d}}_{t} ; \sqrt{\bar{\alpha}_{t}} \textbf{\textit{m}}_{0}, (1-\bar{\alpha}_{t}) \textit{\textbf{I}}\right)
\end{equation}
where $\bar{\alpha}_{t}$ is a constant within the range $(0,1)$ that is monotonically decreasing. As $\mathit{t}$ increases and $\bar{a}_{t}$ approaches $0$, the distribution of $\textbf{\textit{d}}_{t}$ converges to the standard normal distribution $\mathcal{N}(0,\textit{\textbf{I}})$. We use $\textit{T}=1000$ timesteps in our model.

In the denoising process, we follow EDGE~\cite{tseng2023edge} and develop an attention-based network $f_{rev}$ to reverse the forward diffusion process.
The music features $C_{M}$ and text features $C_{E}$ as used as the input conditions to predict the movement of the dance for all $t$. 
We adopt the loss function $\mathcal{L}_{S}$ in DDPM as our objective, optimizing it by learning to estimate $f_{rev} (\textbf{\textit{d}}_{t}, t, C_{M}, C_{E}  ) \approx  \textbf{\textit{m}}_{0}$, where the model refines the noisy latent variable to approximate the true data distribution. Therefore, the training objective can be defined as:
\begin{equation}  
\label{equ_2}
\mathcal{L}_{\text {S}}=
\mathbb{E}_{\boldsymbol{ \textbf{\textit{m}}_{0}}, t}\left[\left\| \textit{\textbf{m}}_{0}- f_{rev} (\textit{\textbf{d}}_{t}, t, 
C_{M}, C_{E}  )\right\|_{2}^{2}\right]
\end{equation}

\subsection{The Proposed DGFM Method}
\begin{figure}[!t]
  \centering
  \includegraphics[width=.9\textwidth]{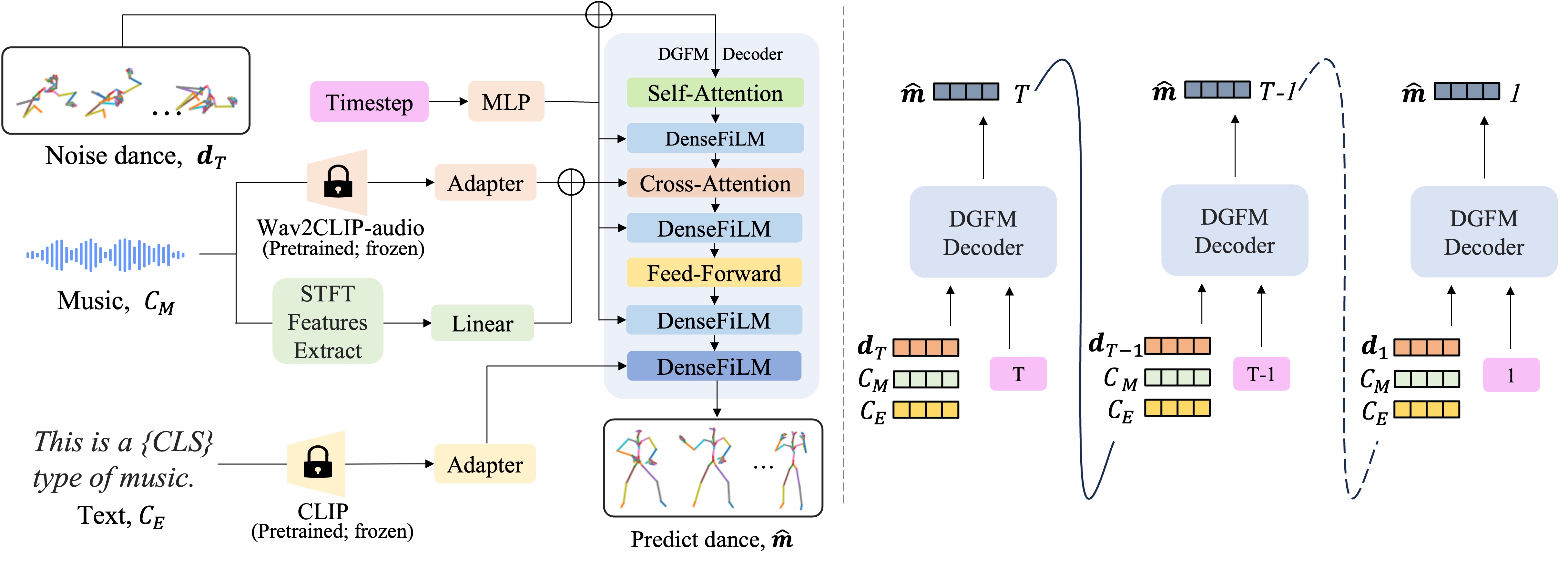}
  \caption{An overview of the proposed DGFM method.}
  \label{fig_1}
  \vspace{-.5cm}
\end{figure}

In this section, we introduce DGFM, a diffusion-based model that utilizes both text and audio inputs to generate full body dance movements while clearly reflecting a distinct dance genre.
The overall architecture of DGDM is illustrated in Figure~\ref{fig_1}. The input music is first divided into $N$ $4$-second segments $\left \{ \hat{C}_{M}^{i}  \right \}_{i=1}^{N}$. 
Next, we extract the $512$-dimensional Wav2CLIP features $\hat{C}_{FM}^{i} \in \mathbb{R}^{T \times 512}$ and the $193$-dimensional STFT features $\hat{C}_{STFT}^{i} \in \mathbb{R}^{T \times 193}$, and feed them into an adapter and a linear layer, respectively.
The linear layer projects the STFT features to a $512$-dimensional vector, then we combine them through the addition operation to obtain $C_{M}\in \mathbb{R}_{}^{T \times 512}$. For the input music genre label, we apply CLIP to extract $512$-dimensional features $C_{E}\in \mathbb{R}_{}^{512}$.

Following EDGE, the input pose data are represented as $ \textbf{\textit{m}} \in \mathbb{R}^{k \times 319}$, according to the Skinned Multi-Person Linear (SMPL) format~\cite{loper2023smpl}. 
This representation consists of three components: a $4$-dimensional foot-ground contact binary label, a $3$-dimensional root translation, and $312$-dimensional rotation information using a $6$-dimensional rotation representation. 
At each denoising timestep $t$, DGFM predicts the denoised result and reintroduces noise from timestep $t-1$ down to $0$.
We repeat this process until all poses are generated at timestep $0$. 
We feed the motion and music features into a Transformer-based denoising network, which consists of a self-attention module, a cross-attention module, multilayer perceptrons, and time-embedding Feature-wise Linear Modulation (FiLM)~\cite{perez2018film}. 
Subsequently, a text-specific FiLM module is incorporated, taking the output from the previous layer $Y$ and the text embedding $C_{E}$ as inputs.
The embeddings are processed as follows:
\begin{equation}  
FiLM_{t}(Y) = \gamma Y + \varepsilon,\quad\gamma  =\theta_{w} (\alpha(C_{E})),\quad\varepsilon  = \theta_{b} (\alpha(C_{E})) 
\end{equation}  
where  $\alpha$ is a text embedding adapter used to adjust the embedding representation. $\theta_{w}$ and $ \theta_{b}$ represent the linear projections responsible for computing the weights and biases, respectively. Apart from the reconstruction loss in Equ.~\ref{equ_2}, we incorporate several auxiliary losses frequently used in motion generation tasks to improve the training stability and physical fidelity. 
Similarly to previous studies~\cite{tevethuman}, we use the forward kinematic function $FK(\cdot)$ to transform the joint angles into their corresponding joint positions, calculating the joint loss 
$ \mathcal{L}_{\text {J}}=\frac{1}{k} \sum_{j=1}^{k}\left\|F K\left(\boldsymbol{\textbf{\textit{m}}}^{j}\right)-F K\left(\hat{\boldsymbol{\textbf{\textit{m}}}}^{j}\right)\right\|_{2}^{2}$,
where $j$ represents the frame index and $\hat{\boldsymbol{\textbf{\textit{m}}}}^{j}$ represents the predicted pose for this frame.
We also compute velocity and acceleration, introducing the velocity loss:
$
\mathcal{L}_{\text {V}}=\frac{1}{k-1} \sum_{j=1}^{k-1}\left\|\left(\boldsymbol{\textbf{\textit{m}}}^{j+1}-\boldsymbol{\textbf{\textit{m}}}^{j}\right)-\left(\hat{\boldsymbol{\textbf{\textit{m}}}}^{j+1}-\hat{\boldsymbol{\textbf{\textit{m}}}}^{j}\right)\right\|_{2}^{2}
$.
Last, we apply the contact loss $\mathcal{L}_{\text {C}}$ which leverages binary foot-ground contact labels to optimize the consistency in foot contact during motion generation:
$
\mathcal{L}_{\text {C}}=\frac{1}{k-1} \sum_{j=1}^{k-1}\left\| \left( F K\left(\hat{\boldsymbol{\textbf{\textit{m}}}}^{j+1}\right)-F K\left(\hat{\boldsymbol{\textbf{\textit{m}}}}^{j}\right) \right) \cdot \hat{\textbf{\textit{b}}}^{j} \right\|_{2}^{2}
$,
where $\hat{\textbf{\textit{b}}}^{j}$ denotes the predicted binary foot-ground contact labels. 
The overall training loss is defined by the weighted sum of the above loss functions $\mathcal{L}=\mathcal{L}_{\text {S}}+\lambda_{\text {J }} \mathcal{L}_{\text {J }}+\lambda_{\text {V }} \mathcal{L}_{\text {V }}+\lambda_{\text {C }} \mathcal{L}_{\text {C }}$, where $\lambda$ are balancing parameters.

\section{Experimental Setup and Results}
\label{others}
\textbf{Dataset:} We evaluate the proposed method on FineDance dataset~\cite{li2023finedance}, which contains $7.7$ hours of paired music and dance, totaling $831,600$ frames at $30$ fps across different $16$ genres. The average dance length is $152.3$ seconds. 
The skeletal data of FineDance is stored in a 3D space and is represented by the standard $52$ joints, including the finger joints.
For all methods, we train on $183$ pieces of music from the training set and generate $270$ dance clips across $18$ songs from the test set, using the corresponding real dances as ground truth.

\textbf{Implementation Details:} The training for our proposed approach uses two NVIDIA GeForce RTX 3090 GPU cards. 
The batch size and total number of epochs are set to $512$ and $2000$, respectively. 
The learning rate is set to $1e-4$, the hidden dimension is set to $512$, and the guidance weight is set to $2.7$.
Four evaluation metrics are used in this paper, including Frechet inception distance (FID)~\cite{siyao2022bailando}, diversity~\cite{siyao2022bailando}, Physical Foot Contact (PFC)~\cite{tseng2023edge} and Beat Alignment Score (BAS)~\cite{onuma2008fmdistance}.

%\textbf{Evaluation Metrics:} We use four evaluation metrics. (1) \textbf{Frechet inception distance (FID)}. Following the approach in~\cite{siyao2022bailando}, we extract the kinetic features~\cite{onuma2008fmdistance} and geometric features~\cite{muller2005efficient} from both the ground truth and generated motions, and use these features to compute the FID scores separately for hands and body. 
%(2) \textbf{Diversity}~\cite{siyao2022bailando} is computed by the average Euclidean distance within the motion features to assess the motion diversity of the body and hands separately.  
%(3) The \textbf{Physical Foot Contact (PFC)}~\cite{tseng2023edge} score evaluates physical plausibility by averaging COM accelerations based on foot stability and contact. 
%(4) The \textbf{Beat Alignment Score (BAS)}~\cite{onuma2008fmdistance} is applied to evaluate beat consistency by calculating the average temporal distance between each music beat and its closest dance beat.

\begin{figure}[!t]
  \centering
  \includegraphics[width=.85\textwidth]{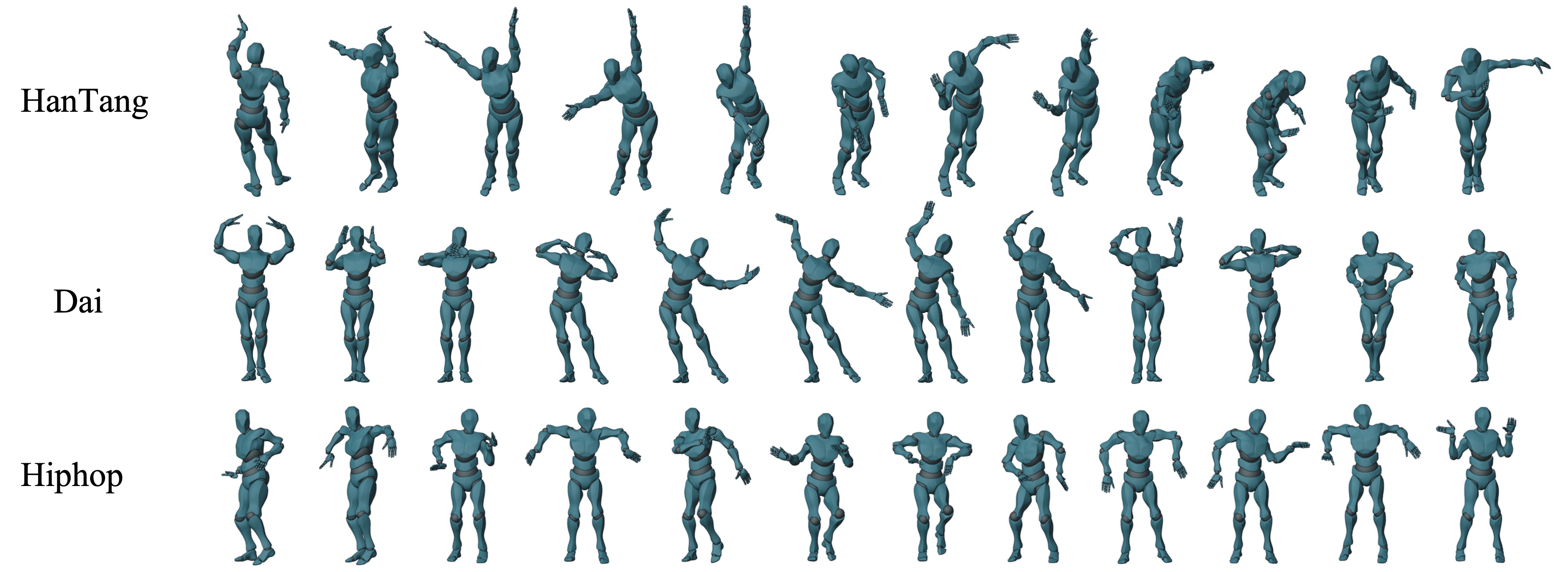}
  \caption{The generated dance motions for three different genres.}
  \label{fig_2}
  \vspace{-.2cm}
\end{figure}

\textbf{Main Results:}
We first visualize the generated dance motions for three different music genres in Figure~\ref{fig_2}. The results are accompanied by the corresponding music, demonstrating that the proposed method is capable of generating realistic and complex dance movements with distinct characteristics. Then we compare the proposed method with four different music foundation models: CLAP~\cite{wu2023large}, Wav2Vec~\cite{baevski2020wav2vec}, Jukebox~\cite{dhariwal2020jukebox} (used by the baseline model EDGE), and Wav2CLIP~\cite{wu2022wav2clip}, as well as two sets of hand-crafted music features: STFT and $35$-D Feature Set, which comprises $35$-dimensional features provided by the FineDance dataset~\cite{li2023finedance}.
The results are reported in Table~\ref{table_1}.

\begin{table}[h!]
\caption{A comparison of different music features. }
\label{sample-table}
\footnotesize
\begin{tabular}{ccccccc}
\toprule
& \multicolumn{2}{c}{Motion Quality} & \multicolumn{2}{c}{Motion Diversity} & \multirow{2}{*}{PFC$\downarrow$} & \multirow{2}{*}{BAS$\uparrow$} \\ \cmidrule(lr{0.35em}){2-3}  \cmidrule(lr{0.35em}){4-5}
& FID\_hand$\downarrow$     & FID\_body$\downarrow$    & Div\_body$\uparrow$      & Div\_hand $\uparrow$    &                         &                                \\ 
\midrule
CLAP~\cite{wu2023large}          & 36.316          & 65.781            & 6.626           & 8.071            & 0.369                   & \underline{0.2256}                   \\
Wav2Vec~\cite{baevski2020wav2vec}       & 35.493       & 30.601                  & 6.508           & 7.257            & 0.241                   & 0.2226                         \\
Jukebox~\cite{dhariwal2020jukebox}       & 38.578          & 53.336                     & 5.687            & 7.453              & 0.223                   & 0.2161                         \\
Wav2CLIP~\cite{wu2022wav2clip}      & 27.212          & 30.539                     & 6.683           & 7.396            & \textbf{0.171}          & \textbf{0.2283}                \\
\hline
STFT          & \underline{18.497}    & \underline {26.522}               & \underline {7.513}      & \textbf{8.562}     & 0.246                   & 0.2201                         \\
35-D Feature Set~\cite{li2023finedance}           & 21.681          & 27.092                     & 7.458            & 8.022             & 0.262                 & 0.2156                         \\
\hline
Wav2CLIP+STFT & \textbf{17.871} & \textbf{25.1752}           & \textbf{7.758}            & \underline {8.2543}      & \underline{0.209}             & 0.2218                                                  \\ 
\bottomrule
\end{tabular}
\label{table_1}
\vspace{-0.3cm}
\end{table}

The experimental results indicate that the dance movements generated using music foundation models generally exhibit better physical realism and consistency with the music. 
Specifically, Wav2CLIP scores $0.171$ in terms of PFC, while Jukebox achieves $0.223$, attributed to their ability to capture a wide range of deep musical features.
In contrast, the dances generated using hand-crafted features show better FID scores and diversity.  
For example, STFT produces FID scores of $18.497$ for hands and $26.522$ for the body, along with diversity scores of $7.513$ and $8.562$ for body and hands.
These results suggest a closer match to the ground truth, although with diminished physical plausibility, as indicated by a PFC score of $0.241$. 
The use of Wav2CLIP and STFT jointly produces the best results, combining the best FID score and diversity for the body, with other metrics also nearly approaching the best results. While music foundation models handle complex musical patterns well, this demonstrates that incorporating hand-crafted features improves generation of dance movements. 

\begin{table}[h]
\caption{A comparison with the state-of-the-art methods. }
\label{sample-table}
\centering
\footnotesize
\begin{tabular}{ccccccc}
\toprule
& \multicolumn{2}{c}{Motion Quality} & \multicolumn{2}{c}{Motion Diversity} & \multirow{2}{*}{PFC$\downarrow$} & \multirow{2}{*}{BAS$\uparrow$} \\ \cmidrule(lr{0.35em}){2-3}  \cmidrule(lr{0.35em}){4-5}
& FID\_hand$\downarrow$     & FID\_body$\downarrow$    & Div\_body$\uparrow$      & Div\_hand $\uparrow$    &                         &                                \\ 
\midrule
DanceRevolution~\cite{huang2020dance}    & 219.312           & 98.402             & 6.773             & 1.813              & 4.199                  & 0.2171                         \\
Bailando~\cite{siyao2022bailando}          & 45.083             & 52.373             & 4.943                & 5.629                & 0.361                   & 0.2152                       \\
EDGE~\cite{tseng2023edge}                 & 38.578               & 53.336               & 5.687                & 7.451                & 0.223                   & 0.2161                         \\
DGFM   & \textbf{20.417}      & \textbf{23.648}      & \textbf{7.631}       & \textbf{8.10}        & \textbf{0.207}          & \textbf{0.2204}                \\
\bottomrule

\end{tabular}
\label{table_2}
\end{table}

We also compare our method with the state-of-the-art dance generation models, including DanceRevolution~\cite{huang2020dance}, Bailando~\cite{siyao2022bailando}, and EDGE~\cite{tseng2023edge}. 
As shown in Table~\ref{table_2}, the proposed method achieves the best results across all the evaluation metrics. 
This highlights the advantages of combining features extracted from Wav2CLIP and STFT. 
The resulting movements not only exhibit high quality and diversity, but also maintain physical and rhythmic fidelity to the music.

\section{Conclusion}
In this paper, we have presented a new diffusion-based 3D dance generation approach using both music and text features. 
To investigate the impact of different foundation music models and hand-crafted music features on the motion generation task, we conducted extensive experiments. 
Specifically, we compared the generation results using four different foundation music models and two sets of hand-crafted music features. 
The experimental results demonstrated that the fusion of Wav2CLIP and STFT features achieves the best performance in terms of motion quality and synchronization with the music. 
Furthermore, we compared our model with several state-of-the-art models, and our model consistently outperformed them in multiple evaluation metrics. 

\newpage

\bibliographystyle{IEEEtran}
\bibliography{neurips_2024}

\end{document}